%% file: main-cvpr26.tex
\documentclass[10pt,twocolumn,letterpaper]{article}

\usepackage{cvpr}  


\usepackage[utf8]{inputenc} 
\usepackage{url}            
\usepackage{booktabs}       
\usepackage{amsfonts}       
\usepackage{nicefrac}       
\usepackage{microtype}      
\usepackage{colortbl}
\usepackage{graphicx}
\usepackage{xspace}
\usepackage{multirow}
\usepackage{tikz}
\usepackage{comment}
\usepackage{enumitem}
\usepackage{makecell}
\usepackage{subcaption}
\usepackage{array}
\usepackage{tcolorbox}
\usepackage{booktabs}
\usepackage[normalem]{ulem}
\useunder{\uline}{\ul}{}

\newcommand{\sys}{{MobileViews}\xspace}

\newcommand{\deltapos}[1]{{\scriptsize\textcolor{blue}{$\uparrow$#1}}}

\usepackage{pifont}
\newcommand{\xmark}{\textcolor{red}{\ding{55}}\xspace}
\newcommand{\cmark}{\textcolor{blue}{\ding{51}}\xspace}
\definecolor{lightpink}{rgb}{1, 0.8, 0.8}

\definecolor{cvprblue}{rgb}{0.21,0.49,0.74}
\usepackage[pagebackref,breaklinks,colorlinks,allcolors=cvprblue]{hyperref}

\title{\sys: A Million-scale and Diverse Mobile GUI Dataset}

\author{
Longxi Gao$^{1}$\thanks{Equal contribution.},
Li Zhang$^{1}$\footnotemark[1],
Shihe Wang$^{1}$,
Pengzhi Gao$^{3}$,\\
Wei Liu$^{3}$,
Jian Luan$^{3}$,
Shangguang Wang$^{1}$,
Yuanchun Li$^{2}$,
Mengwei Xu$^{1}$\\[4pt]
$^{1}$Beijing University of Posts and Telecommunications\\
$^{2}$Institute for AI Industry Research (AIR), Tsinghua University\\
$^{3}$Unaffiliated\\
{\tt\small glx@bupt.edu.cn}
}

\begin{document}
\maketitle
\input{sections/sec-abstract.tex}
\input{sections/sec-intro.tex}
\input{sections/sec-related-work.tex}
\input{sections/sec-construction.tex}
\input{sections/sec-dataset-details.tex}

\input{sections/sec-experiment.tex}
\input{sections/sec-conclusion.tex}
{
    \small
    \bibliographystyle{ieeenat_fullname}
    \bibliography{agents-ref}
}

\end{document}

%% file: sections/sec-abstract.tex
\begin{abstract}
    Visual language models (VLMs) empower mobile GUI agents to interpret complex mobile screens and respond to user requests.
    Training such capable agents requires large-scale, high-quality mobile GUI data.
    However, existing mobile GUI datasets are limited in scale, data comprehensiveness, and fidelity.
    To overcome this, we utilize two mobile SoC clusters to provide over 200 native, high-fidelity mobile environments, along with a VLM-enhanced automatic application traversal framework for highly parallel, automated dataset collection with minimal human intervention.
    With this system, we propose \sys, a million-scale mobile GUI dataset comprising over 1.2 million unique screenshot-view hierarchy pairs from more than 30K modern Android applications.
    We assess the effectiveness of \sys by training four VLMs using the reinforcement learning-based GUI grounding task and evaluating them on two representative GUI grounding benchmarks.
    Results show that \sys significantly enhances grounding accuracy by up to 6.1\%.
    Further analysis of data scale and quality underscores the critical role of large, high-quality datasets as reliable sources for training mobile GUI agents.
    The \sys dataset is publicly available at \url{https://huggingface.co/datasets/mllmTeam/MobileViews}.
\end{abstract}

%% file: sections/sec-intro.tex
\section{Introduction}

\begin{figure}[t]
    \centering
    \includegraphics[width=0.48\textwidth]{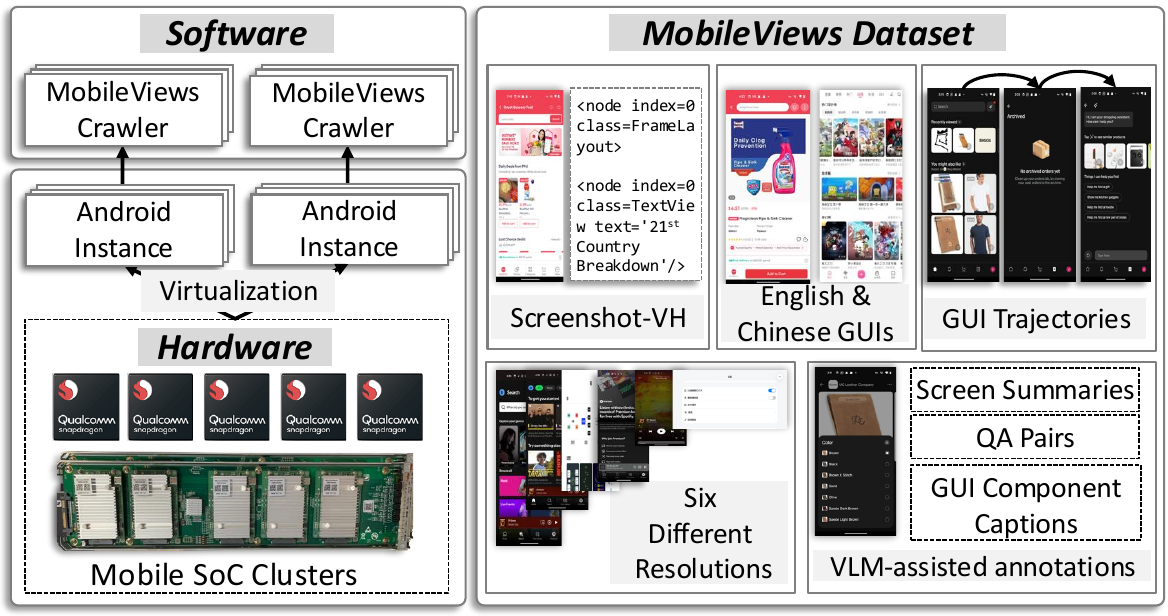}
    \caption{Overview of \sys and its collection infrastructure.
    We employ the \sys Crawler to automatically interact with native Android instances, which are powered by Android virtualization-enhanced Mobile SoC Clusters.}
    \label{fig:teaser}
    \vspace{-13pt}
\end{figure}

Visual language models (VLMs) empower mobile GUI agents to understand screens and respond to diverse user requests, benefiting individuals with disabilities and hands-free scenarios such as driving~\cite{hong2024cogagent,yang2023appagent,lin2024showui}.
During pre-training, general-purpose VLMs gain basic GUI understanding and reasoning abilities by incorporating a small portion of GUI data within billion-scale natural image-text pairs~\cite{bai2025qwen25vltechnicalreport,coreteam2025mimovltechnicalreport,chen2023internvl}.
However, natural images depicting real-world scenes still dominate the training corpus, while mobile screens with dense layouts, structured components, and text-rich content remain severely underrepresented.
This imbalance limits the models' ability to handle complex GUI tasks.
Therefore, large-scale and comprehensive GUI data may empower general-purpose VLMs to handle GUI-specific tasks.

\input{tables/dataset-comparison.tex}
\textbf{Existing datasets and limitations.}
As shown in Table~\ref{tab:ui-dataset-comp}, existing open-source mobile GUI datasets face a critical trade-off between scale and comprehensiveness.
Rico~\cite{deka2017rico}, proposed in 2017, includes only 63K unique screens at a single screen resolution.
While AITW~\cite{rawles2023android} contains 2.28M unique screens, it suffers from two key limitations: (1) limited app diversity results in high content similarity (as we will shown in $\S$\ref{sec:dataset-preview}), and (2) the lack of view hierarchies (VHs) hinders fine-grained and accurate screen interpretation in downstream training tasks.
GUIOdyssey~\cite{lu2025guiodyssey} and AndroidControl~\cite{li2024effectsdatascaleui} provide rich textual annotations but rely on manual construction, leading to prohibitive costs and limited scalability.
Additionally, most datasets contain only English GUIs and single-resolution screens, lacking the linguistic and visual diversity needed for generalized GUI agents.
These limitations call for a large-scale and comprehensive mobile GUI dataset built upon a scalable collection pipeline.

\textbf{Challenges and key techniques.}
Collecting large-scale mobile GUI datasets faces fundamental challenges due to inefficiencies at both the software and hardware levels.

$\bullet$
At the software level, efficient and scalable traversal across diverse mobile apps is essential, yet current solutions do not fully address this requirement.
(1) Static rule-based GUI navigation tools~\cite{li2017droidbot,android-monkey} often yield low-quality screen data, as they struggle to adapt to complex GUI contexts and diverse navigation patterns.
For instance, they frequently get stuck on login pages or intrusive pop-ups.
(2) Most existing large mobile GUI datasets~\cite{deka2017rico,rawles2023android,li2024effectsdatascaleui,lu2025guiodyssey} are manually collected to ensure high quality, as shown in Table~\ref{tab:ui-dataset-comp}. 
However, such manual collection process is not scalable due to the substantial human labor required.

\textbf{Design\#1: VLM-enhanced automatic app crawler.}
To ensure both efficiency and data quality, we design \sys Crawler, a VLM-enhanced automatic traversal framework that follows predefined interaction rules and invokes VLM to process a list of pre-defined triggers. 
Specifically, the crawler captures the VH metadata, simplifies it, identifies and traverses interactable GUI components sequentially.
When predefined triggers occur, such as prolonged idleness, login pages, or pop-ups, the VLM is activated to assist further traversal.
This hybrid strategy maximizes GUI coverage and minimizes manual intervention, which is required only when both predefined rules and the VLM fail to proceed.

$\bullet$ At the hardware level, large-scale app traversal requires running numerous Android instances in parallel with high environment fidelity. 
Most mobile GUI datasets rely on emulators~\cite{toyama2021androidenv,sun2022meta,rawles2023android,zhang2024llamatouch} for cost-effective scaling.
However, emulators typically run on x86 machines, causing compatibility issues with ARM-based apps and limiting app diversity.
For instance, AITW~\cite{rawles2023android} collects about 5M screens from only 357 apps on Android emulators, but fewer than 2.3M are left after imaeg de-duplication.
In contrast, Rico~\cite{deka2017rico} uses physical device farms to run apps natively, enabling collection from over 9K apps.
Although physical devices eliminate compatibility constraints, they face scalability challenges due to high operational costs and space constraints.

\textbf{Design\#2: Scalable, high-fidelity app traversal on SoC clusters.}
To improve hardware-level traversal efficiency, we deploy two SoC clusters~\cite{zhang2024soccluster} with 120 Snapdragon processors~\cite{snapdragon}, which provide native Android environments and high app compatibility.
On top of this, we leverage Android OS virtualization to launch over 200 instances across the two SoC clusters.
Together with fine-grained task scheduling, this setup achieves high parallelism and significantly reduces data collection time, which gathers over 1.2M unique screens in approximately 151,600 device-hours.

\textbf{The new dataset: \sys.}
Based on the proposed system, we construct \sys, an open-source dataset with over 1.2M unique screenshot-VH pairs from more than 30K Android apps.
Compared to AndroidControl~\cite{li2024effectsdatascaleui}, \sys includes 36$\times$ more apps and 15$\times$ more screenshot-VH pairs, with a scalable pipeline for flexible expansion.
Beyond scale, \sys covers six screen resolutions, both English and Chinese GUIs, and complete GUI trajectories.
We further enhance \sys with 13K screen summaries, 26K QA pairs, and 26K GUI component captions.
Detailed dataset statistics are presented in $\S$\ref{sec:dataset-preview}.

\begin{figure*}[t]
	\centering					
	\includegraphics[width=0.85\textwidth]{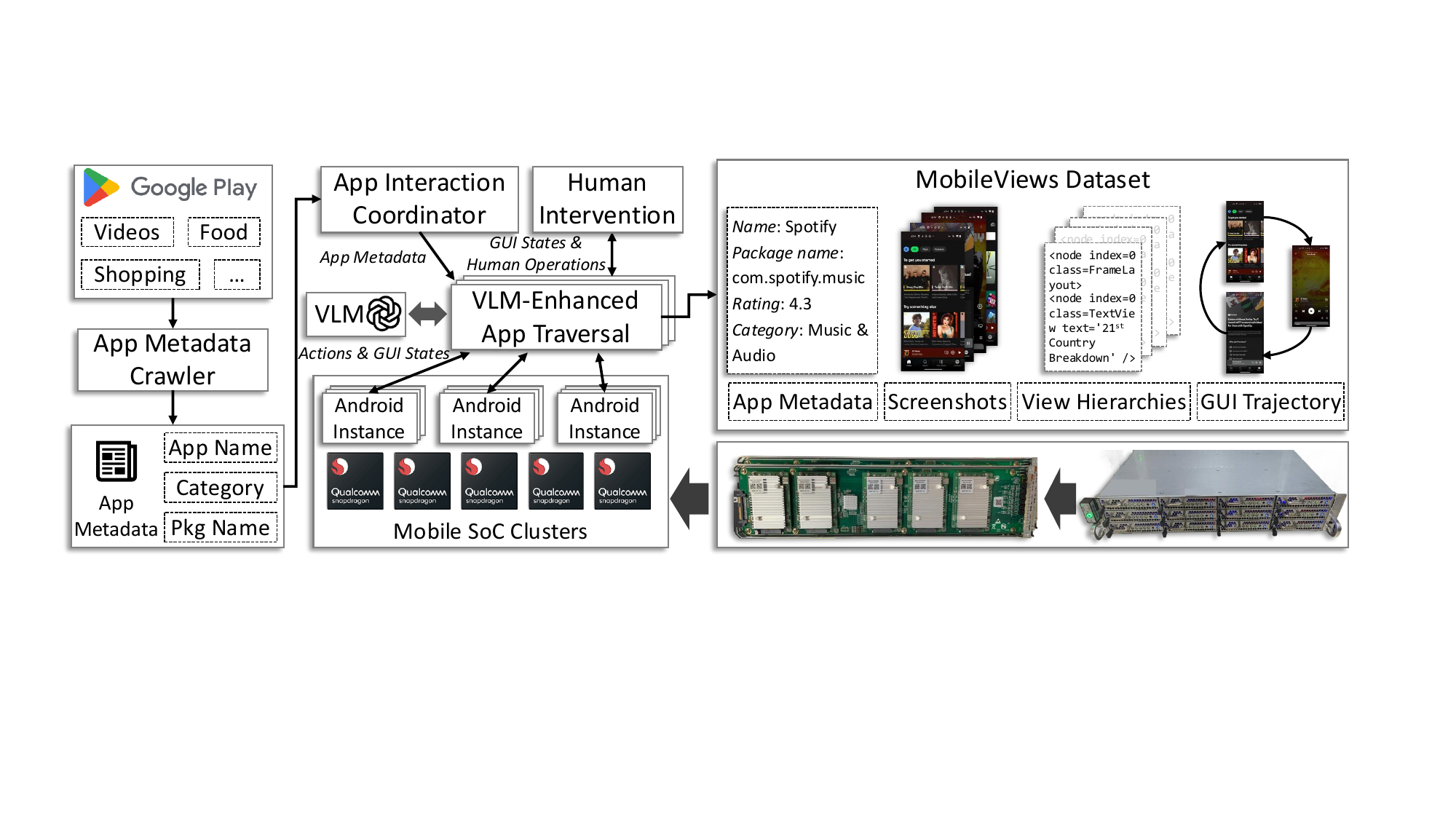}
	\caption{Left: Workflow of the \sys dataset construction. Right: Demonstration of the \sys dataset, and a physical SoC Cluster to empower high parallelism app traversal.
    }
    \vspace{-10pt}
	\label{fig:dataset-construction}
\end{figure*}

\textbf{Experiments.}
We evaluate \sys through GUI grounding experiments on four VLMs: three general-purpose models (Qwen2.5-VL-7B~\cite{bai2025qwen25vltechnicalreport}, Mimo-VL-7B-SFT~\cite{coreteam2025mimovltechnicalreport}, Mimo-VL-7B-RL~\cite{coreteam2025mimovltechnicalreport}) and one GUI-specific model (UI-Tars-1.5~\cite{qin2025uitars}).
All models are reinforcement fine-tuned on the annotated subset and evaluated on two GUI grounding benchmarks, ScreenSpot-v2~\cite{wu2024atlas} and ScreenSpot-Pro~\cite{li2025screenspotproguigroundingprofessional}.
Results demonstrate that training with \sys consistently enhances GUI grounding performance across all models.
After fine-tuning, UI-Tars-1.5 achieves 91.2\% on ScreenSpot-v2 and 47.7\% on ScreenSpot-Pro, with improvements of 2.2\% and 6.1\%, respectively.
With an additional training data filtering strategy applied, Mimo-VL-7B-SFT improves from 87.6\% to 91.0\% on ScreenSpot-v2, while Mimo-VL-7B-RL improves from 40.2\% to 43.5\% on ScreenSpot-Pro.
These findings confirm that \sys serves as a large-scale, high-quality dataset for improving both general-purpose and GUI-specific VLMs.

\textbf{Contributions} of this study are as follows:
\begin{itemize}
    \item A framework with a joint software-hardware design for high-fidelity, automated mobile GUI data collection.
    \item \sys, a million-scale mobile GUI dataset comprising over 1.2M unique screens from more than 30K Android apps, will be fully open-sourced.
    \item Experiments on four VLMs demonstrate that training with \sys improves GUI grounding performance across models by up to 6.1\%.
\end{itemize}

%% file: tables/dataset-comparison.tex
\begin{table*}[t]
    \vspace{-3pt}
    \centering	
    \caption{Comparison of representative open-source mobile GUI datasets. For large datasets (i.e., those with more than 50,000 screens), we count the number of unique screens using ImageHash~\cite{imagehash} to eliminate duplicates.}
    \vspace{-3pt}
    \scalebox{0.73}{
    \begin{tabular}{|c|cc|ccccc|cc|}
    \hline
    \multirow{2}{*}{\textbf{\begin{tabular}[c]{@{}c@{}}\\Mobile GUI Dataset\end{tabular}}} 
    & \multicolumn{2}{c|}{\textbf{Scale}}                                            & \multicolumn{5}{c|}{\textbf{Data Comprehensiveness}}                                                                                                                        & \multicolumn{2}{c|}{\textbf{Data Collection}}                                                             \\ \cline{2-10}
                                                                                                & \textbf{Apps}   & \textbf{\begin{tabular}[c]{@{}c@{}}Unique\\ Screens\end{tabular}}   & \textbf{\begin{tabular}[c]{@{}c@{}}App\\ Metadata\end{tabular}} & \textbf{\begin{tabular}[c]{@{}c@{}}Screenshot-\\ VH Pairs\end{tabular}}  & \textbf{\begin{tabular}[c]{@{}c@{}}GUI\\ Trajectories\end{tabular}} & \textbf{\begin{tabular}[c]{@{}c@{}}Multiple\\ Resolutions\end{tabular}} & \textbf{\begin{tabular}[c]{@{}c@{}}Languages\\ (EN / ZH)\end{tabular}} & \textbf{Automation}                                                                 & \textbf{Hardware}     \\ \hline
    Rico~\cite{deka2017rico}                                                                                        & 9,772           & 63,370             & \cmark                                                                              & \cmark          & \cmark                                                              & \xmark                                                     & EN                                                         & \xmark                                                                  & Physical Devices \\
    PixelHelp~\cite{li2020mapping}                                                                                   & 4               & 187                & \xmark                                                                               & \cmark          & \cmark                                                              & \xmark                                                     & EN                                                           & \xmark                                                                  & Emulators      \\
    Screen2words~\cite{wang2021screen2words}                                                                                & 6,269           & 22,417             & \cmark                                                                              & \cmark          & \xmark                                                               & \xmark                                                     & EN                                                           & \xmark                                                                  & N/A                   \\
    ScreenQA~\cite{baechler2024screenai}                                                                                    & N/A             & 35,352             & \cmark                                                                              & \cmark          & \xmark                                                               & \xmark                                                     & EN                                                           & \xmark                                                                  & N/A                   \\
    META-GUI~\cite{sun2022meta}                                                                                    & 11              & 24,825             & \xmark                                                                               & \cmark          & \cmark                                                              & \xmark                                                     & EN                                                           & \xmark                                                                  & Physical Devices  \\
    DroidTask~\cite{wen2024autodroid}                                                                                   & 13              & 362                & \xmark                                                                              & \cmark          & \cmark                                                              & \xmark                                                     & EN                                                           & \xmark                                                                  & Emulators      \\
    AITW~\cite{rawles2023android}                                                                                        & 357            & 2,282,533          & \xmark                                                                              & \xmark           & \cmark                                                              & \cmark                                                     & EN                                                           & \xmark                                                                  & Emulators      \\
    LlamaTouch~\cite{zhang2024llamatouch}                                                                                  & 57              & 3,281              & \xmark                                                                              & \cmark          & \cmark                                                              & \xmark                                                     & EN                                                           & \xmark                                                                  & Emulators      \\ 
    GUIOdyssey~\cite{lu2025guiodyssey}                                                                                  & 212              & 77,535              & \xmark                                                                              & \xmark          & \cmark                                                              & \cmark                                                     & EN                                                           & \xmark                                                                  & Emulators      \\ 
    AndroidControl~\cite{li2024effectsdatascaleui}                                                                                  & 833              & 81,459              & \xmark                                                                              & \cmark          & \cmark                                                              & \xmark                                                     & EN                                                           & \xmark                                                                  & Physical Devices      \\ 
    GUI-Net-1M~\cite{zhang2025tongui}                                                                                  & 280              & 720,374              & \xmark                                                                              & \xmark          & \cmark                                                              & \cmark                                                     & EN \& ZH                                                           & \xmark                                                                  & N/A      \\ 
    \hline
    \textbf{\sys}                                                                         & \textbf{30,037} & \textbf{1,213,866} & \textbf{\cmark}                                                           & \textbf{\cmark} & \textbf{\cmark}                                                     & \textbf{\cmark}                                           & \textbf{EN \& ZH}                                                 & \textbf{\cmark}                                                     & \textbf{SoC clusters} \\ \hline
    \end{tabular}
    }
    \vspace{-8pt}
    \label{tab:ui-dataset-comp}
\end{table*}

%% file: sections/sec-related-work.tex
\section{Related Work}
\label{sec:bkgnd-screen-datasets}

In this section, we compare existing open-source mobile GUI datasets from three perspectives: (1) data scale, (2) data comprehensiveness, and (3) data collection methods.
Table~\ref{tab:ui-dataset-comp} summarizes these key characteristics.

\textbf{Data scale.}
Rico~\cite{deka2017rico} is a widely-used dataset released in 2017, containing about 63K unique screens collected from over 9K Android apps.
Several studies have augmented Rico with additional annotations, such as Screen2Words~\cite{wang2021screen2words} and ScreenQA~\cite{hsiao2022screenqa}, but remain limited in scale.
Beyond static screen understanding, some datasets~\cite{li2020mapping,rawles2023android,zhang2024llamatouch,li2024effectsdatascaleui,lu2025guiodyssey,wu2024mobilevlm} focus on training and benchmarking GUI grounding or navigation models.
However, most of these datasets are collected from a limited number of apps and contain relatively few screens, yet recent GUI-specific VLMs demonstrate that large-scale, high-quality data is critical for effective GUI understanding.
For instance, UI-Venus~\cite{gu2025uivenus} selects only 107K high-quality samples from 627K raw GUI grounding data to improve GUI grounding performance, showing that effective quality control requires sufficiently large data pools.

\textbf{Data comprehensiveness.}
From an informational perspective, both pixel-level screenshots and textual VHs are essential for mobile GUI understanding.
While screenshots capture the exact visual state of a screen and provide a quick reference for layout and image-based components, VHs record the screen's structural metadata, including component hierarchy, bounding boxes, and attributes that are not directly visible.
Together, these two modalities offer a comprehensive understanding of both the visual and structural aspects of mobile screens.
Although AITW~\cite{rawles2023android} reaches the largest scale to date, about 2.3M screens from 357 apps, it lacks VH metadata, preventing flexible sample construction by combining both visual and structural information. 
Similarly, GUI-Net-1M~\cite{li2024effectsdatascaleui}, which is collected from online videos (e.g., YouTube, Bilibili) instead of direct app traversal, also lacks VHs.
Moreover, most datasets do not account for the diversity of screen resolutions or languages, restricting models' cross-device and cross-language generalization.
Overall, existing datasets are far from comprehensive in capturing both visual and structural aspects of mobile GUIs.

\textbf{Data collection methods.}
The data collection platforms and strategies are both crucial for ensuring the quality and scalability of mobile GUI datasets.
Datasets such as Rico~\cite{deka2017rico} rely on physical devices, which provide high-fidelity environments but face scalability challenges due to cost and space constraints.
AITW~\cite{rawles2023android} and GUIOdyssey~\cite{lu2025guiodyssey} use emulators to collect data, which offers cost-effective scaling but suffers from compatibility issues since most mobile apps are designed for ARM processors, whereas emulators typically run on x86 machines.
This incompatibility restricts app diversity and limits the coverage of modern apps.
Beyond environment constraints, some datasets rely on manual collection, which is prohibitively time-consuming and costly. 
For instance, AndroidControl~\cite{li2024effectsdatascaleui} takes one year to collect 14,548 episodes over 833 Android apps.
These limitations motivate the construction of \sys, which combines VLM-enhanced automatic app traversal with high-fidelity SoC clusters to enable scalable and reliable data collection across diverse modern mobile applications.

%% file: sections/sec-construction.tex
\section{\sys Dataset Construction}

Given the limitations of preivous datasets in terms of scale and comprehensiveness, we construct a new mobile GUI dataset, \sys.
It contains over 1.2M unique screenshot-VH pairs collected from more than 30K modern mobile apps, 
covering six different resolutions, both English and Chinese GUIs, and complete GUI trajectories.
Building such a large-scale dataset poses a key challenge: \textit{how to efficiently collect large-scale, comprehensive mobile screens with minimal human intervention}.
To address this, we propose \sys Crawler, a scalable app traversal framework for automated data collection.
As illustrated in Figure~\ref{fig:dataset-construction}, it includes a \textit{a VLM-enhanced automatic app traversal tool} that processes individual apps on each Android instance by combining fixed interaction rules, VLM, and minimal human intervention to achieve efficient app traversal.
This workflow is further scaled across two SoC clusters to enable high-parallel app exploration.

\subsection{App Traversal Workflow}

\begin{table}[]
    \centering
    \caption{Comparison of three different app traversal methods.
    \textbf{\#Apps} is the number of apps successfully traversed without stuck.
    }
    \vspace{-5pt}
    \scalebox{0.9}{
    \begin{tabular}{c|ccc}
    \toprule
    \textbf{Metrics} & \textbf{\begin{tabular}[c]{@{}c@{}}Fully\\ Rule-based\end{tabular}} & \textbf{\begin{tabular}[c]{@{}c@{}}W/\\ VLM\end{tabular}} & \textbf{\begin{tabular}[c]{@{}c@{}}W/ Human\\ Intervention\end{tabular}} \\ 
    \midrule
    \textbf{\#Apps (Total = 86)}                 & 16                                                                  & 38                                                        & 86                                                                        \\
    \textbf{Success Rate}              & 18.6\%                                                              & 44.2\%                                                    & 100\%                                                                     \\ 
    \bottomrule
    \end{tabular}
    }
    \footnotesize
    \label{tab:app-traversal-rate}
    \vspace{-12pt}
\end{table}

\textbf{App metadata collection.}
The data collection process begins with gathering app metadata, such as app package names.
\sys covers apps from 33 categories on the Google Play Store, including many trending ones, providing a more accurate reflection of recent mobile interface designs.

\textbf{App interaction coordination.}
The collected metadata is sent to an \textit{App Interaction Coordinator}, which serves as the overall traversal task manager.
The coordinator manages Android instances provided by the SoC clusters and initializes each traversal process.
Each process starts with app installation and launching.
Specifically, the coordinator sends an app package name (e.g., \textit{com.spotify.music}) to the traversal tool on an available Android instance.
The traversal tool then searches for this app on the Google Play Store using GUI automation scripts, installs it, and launches it via the Android Debug Bridge~\cite{adb}.
Once launched, the traversal tool begins interacting with the app to collect screenshots and VHs.
If the step limit is reached or an unexpected interruption occurs, the traversal tool stops.
The coordinator will record the current system and app states for later analysis.

\textbf{VLM-enhanced automatic app traversal.}
By default, the traversal tool follows predefined rules to interact sequentially with all interactable GUI components on each screen.
When the traversal gets stuck, such as on login pages, the VLM steps in to interpret complex layouts and guide subsequent interactions to access deeper interfaces.

\subsection{App Interaction with \sys Crawler}
Traditional app traversal methods rely heavily on human effort to ensure data quality.
However, such manual approaches are not scalable when building datasets involving a large number of apps and screens.
Our proposed \sys Crawler, a VLM-enhanced automatic app traversal framework, minimizes human intervention while maintaining both efficiency and scalability.

\begin{figure}[t]
    \centering
    \includegraphics[width=0.48\textwidth]{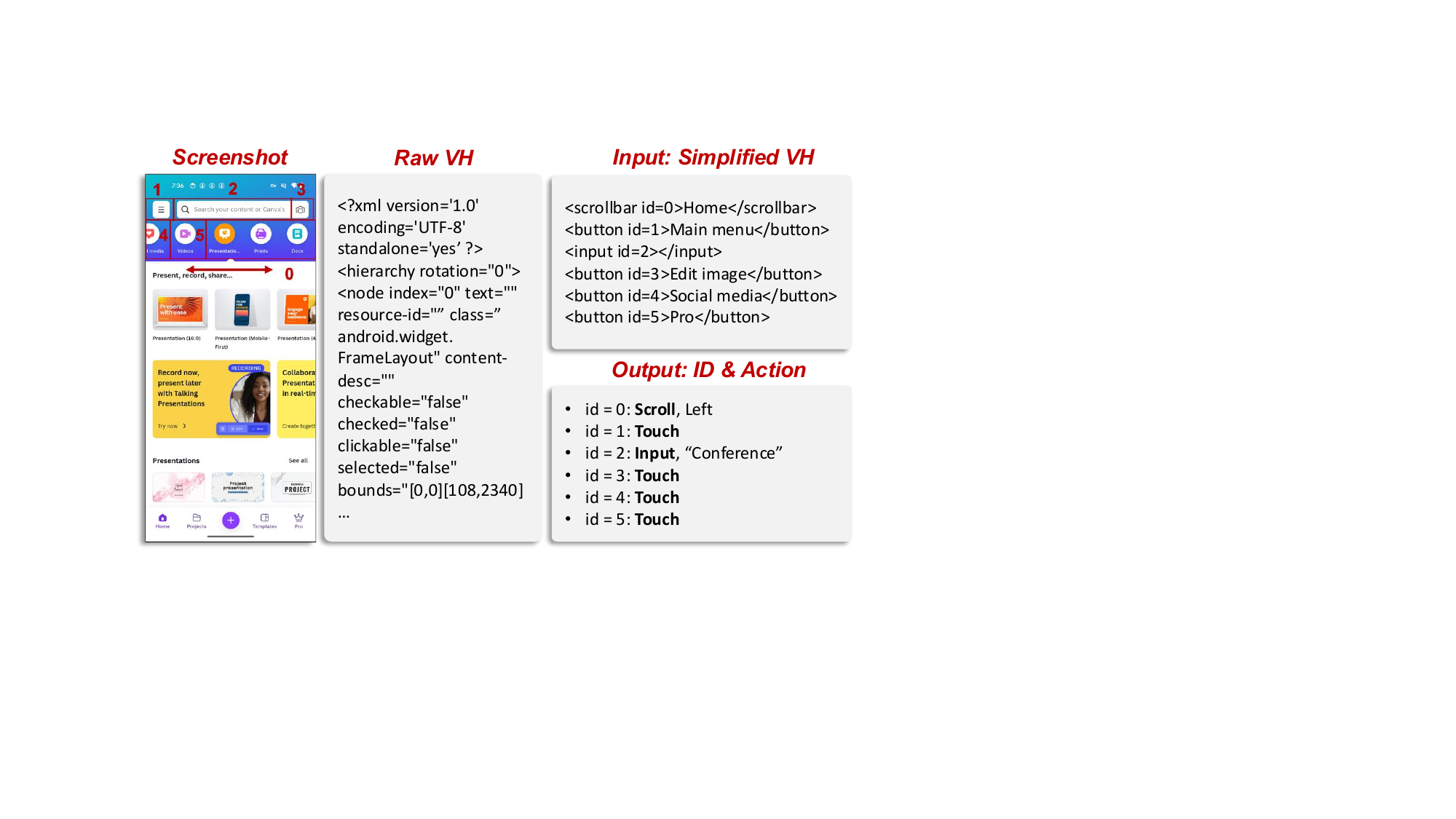}
    \caption{Example of a single screen interaction. The traversal tool simplifies the raw VH and identifies the interactable GUI components along with their corresponding actions.}
    \label{fig:automatic-traversal-demo}
    \vspace{-13pt}
\end{figure}

\textbf{Rule-based app interaction.}
We employ a rule-based app traversal tool, Droidbot~\cite{li2017droidbot}, to automate the traversal process.
As illustrated in Figure~\ref{fig:automatic-traversal-demo}, Droidbot captures the VH of the current screen and identifies interactable GUI components based on specific attributes such as \textit{clickable}, \textit{editable}, and \textit{focusable}. 
It then executes corresponding actions (e.g., clicks, scrolls, or text inputs) on these GUI components in sequence.
When Droidbot navigates to a new screen, it repeats this procedure until reaching the maximum number of steps, which is set to 1,000 actions per app in our setup.

\textbf{VLM integration.}
While fixed interaction rules can reach many common GUIs effectively, they often fail to handle complex ones, such as accurately filling in specific input fields or closing instrusive pop-ups.
As shown in Table~\ref{tab:app-traversal-rate}, the rule-based approach successfully traverses only 18.6\% (16 out of 86 trending apps on the Google Play Store) apps without getting stuck.
This low success rate arises from multiple challenges: many apps require registration or login to access deeper interfaces, intrusive pop-ups block navigation, and unintended links to external apps disrupt the traversal flow.
To address this, we integrate a VLM to handle complex GUI states under two primary conditions:
(1) the traversal tool encounters GUI states matching predefined triggers (e.g., keywords like ``sign up'' or ``login'' are detected in the VH),
and (2) the current screen remains idle beyond a defined timeslot.
In these cases, the VLM takes over the traversal process until it reports completion, either successful or failed.
Utilizing GPT-4o~\cite{gpt4o}, the traversal success rate increases to 44.2\% (38 out of 86 apps).
This VLM-enhanced framework can also be extended to other scenarios by defining suitable triggers to handle various complex GUI states.

\textbf{Human intervention.}
Although the VLM integration can handle complex GUI states effectively, certain challenging cases can still cause the traversal process to stall.
This typically occurs when the VLM reaches its maximum retry limit or returns a ``task failed'' response.
To address such situations, we adopt a \textit{lazy processing} strategy in which human operators intervene only when the traversal tool encounters an unprocessable GUI state.
In these cases, the tool stops exploration and the \textit{App Interaction Coordinator} records the current state.
Human operators later review these apps in batches, manually resolve the blocked states, and resume automatic traversal.
This strategy significantly reduces manual effort while maintaining traversal coverage and efficiency.

\subsection{SoC Cluster-Powered Scalable App Interaction}
By combining rule-based interaction, VLM, and human intervention, the traversal tool can efficiently navigate each app and collect diverse GUIs.
The next challenge lies in scaling this process to handle large number of apps in native mobile environments.
Existing approaches for mobile GUI dataset construction, such as running apps on physical device farms or Android emulators, both have notable limitations.
While physical device farms are costly to scale, Android emulators running on x86-based machines often face compatibility issues with apps built for ARM-based processors.
To ensure both scalability and compatibility, we deploy two SoC clusters~\cite{zhang2024soccluster} equipped with 120 Snapdragon~\cite{snapdragon} processors, which provide native ARM-based execution environments for large-scale mobile app interaction.
Figure~\ref{fig:dataset-construction} shows the physical setup of an SoC cluster and its network-interconnected mobile processors inside the server.
To maximize resource utilization, we further implement SoC-level virtualization to run multiple Android instances on each processor.
As a result, more than 200 Android instances run in parallel across the two SoC clusters, enabling large-scale app interaction in a more efficient and cost-effective manner than traditional device farms.

\textbf{Software-hardware coordination.}
To achieve better parallelism, the \textit{App Interaction Coordinator} monitors the status of each Android instance at regular intervals.
When an instance becomes available, the coordinator connects to it and initiates the app traversal process.
The traversal tool then starts its execution.
Once the traversal is completed or the process exits unexpectedly, the instance is marked as available again.
This fine-grained scheduling strategy enables efficient utilization of hardware resources, allowing us to complete 151,600 device-hours of app traversal within two months.

%% file: sections/sec-dataset-details.tex
\section{\sys Dataset}
\label{sec:dataset-preview}

As shown in Table~\ref{tab:ui-dataset-comp}, \sys contains over 1.2M unique mobile screens collected from more than 30K Android apps, demonstrating clear advantages over existing datasets in scale, comprehensiveness, and collection scalability.
In this section, we detail \sys's composition and analyze its screen diversity.

\textbf{\sys composition and features.}
\sys is a large-scale and diverse resource for training mobile GUI agents with over 1.2M mobile screenshot-VH pairs.
It also provides GUIs across multiple resolutions, languages, and trajectories.
Specifically, screenshots capture the visual state of each mobile interface, offering a direct view of layouts and image-based components.
VHs contain detailed structural metadata describing GUI component hierarchies and attributes, including bounding boxes, component types, and interaction properties, serving as a valuable foundation for constructing diverse training samples.
Multiple resolutions enhance model generalization across diverse device configurations.
For English GUIs, \sys includes five resolution settings: 720$\times$1280, 720$\times$1520, 1080$\times$2340, 1440$\times$2560, and 1920$\times$1080, each containing more than 60K samples, plus an additional 1080$\times$1920 setting covering the remaining English screens.
Chinese GUIs comprise 50K unique screens, all at 1080$\times$1920 resolution, strengthening bilingual support.
GUI trajectories are naturally captured during the automated traversal process, with each trajectory recording the sequence of screens and interactions to provide valuable contextual information for studying GUI navigation patterns.
By integrating all these features at scale, \sys offers high data diversity and broad coverage, making it suitable for training capable mobile GUI agents.

\begin{figure}[t]
    \centering
    \includegraphics[width=0.48\textwidth]{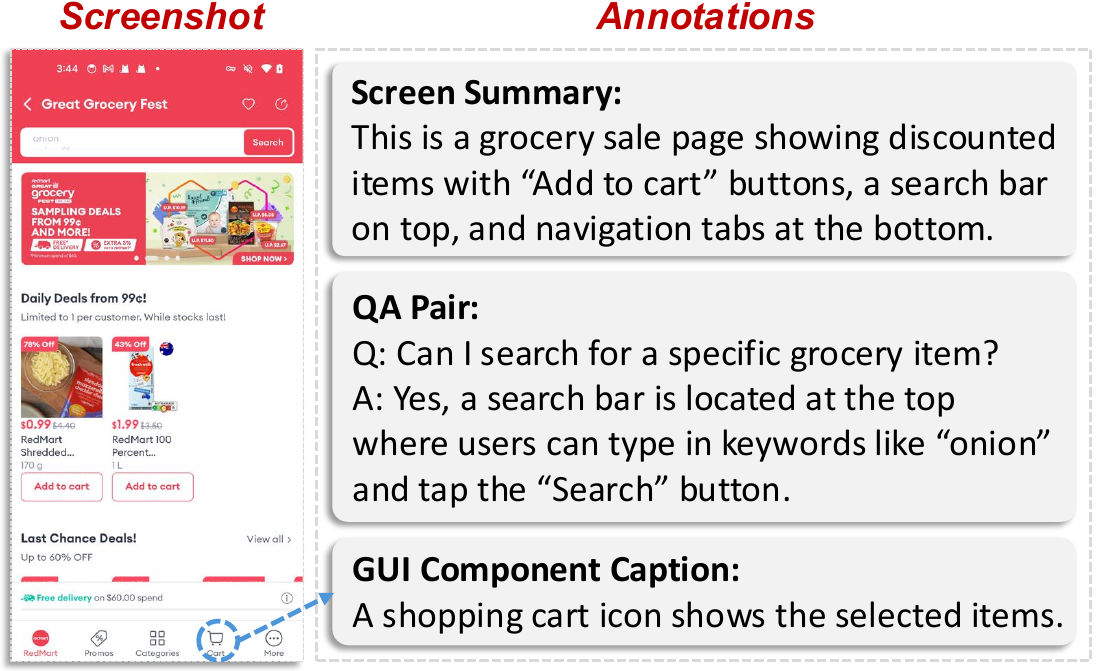}
    \caption{Illustration of VLM-assisted annotations, including a screen summary, a QA pair, and a GUI component caption.}
    \label{fig:annotation-instance}
    \vspace{-11pt}
\end{figure}

\textbf{Enhance \sys with VLM-assisted annotations.}
To support various GUI tasks, we augment \sys with VLM-assisted annotations for both mobile GUI understanding and grounding.
First, we select 20K unique screens from \sys as annotation candidates.
We then extract metadata from each screen's VH, including component interactability, bounding boxes, and text labels, which serve as the basis for multi-stage filtering.
To ensure data diversity and annotation quality, this filtering proceeds in two steps:
\begin{itemize}
    \item \textit{Screen-level filtering.} We exclude screens with fewer than three interactable GUI components to ensure sufficient content richness for meaningful annotations.
    \item \textit{Component-level filtering.} We retain only clickable GUI components with non-empty text descriptions and remove components with duplicate labels on the same screen to reduce annotation ambiguity. We further use OCR to verify that component text is clearly visible and readable, filtering out components with obscured text.
\end{itemize}

After filtering, we obtain 13K representative screens with clear and diverse GUI components.
We then employ GPT-4o~\cite{gpt4o} to generate annotations for each screen: one screen summary describing the overall interface, two QA pairs about screen functionality, and two GUI component captions describing selected GUI components.
This produces an annotated subset of 13K screens with corresponding summaries, 26K QA pairs, and 26K GUI component captions, providing reliable supervision for downstream GUI understanding and grounding tasks.
Figure~\ref{fig:annotation-instance} illustrates examples of the VLM-assisted annotations.

\begin{figure}[t]
    \centering
    \begin{subfigure}[t]{0.2\textwidth}
        \includegraphics[width=\textwidth]{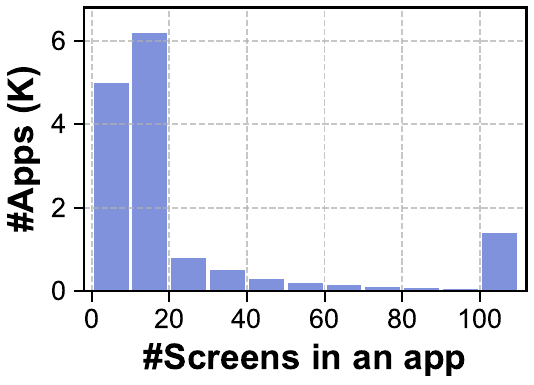}
        \caption{}
        \label{fig:ui-cnt-distribution}
    \end{subfigure}
    \hspace{3pt}
    \begin{subfigure}[t]{0.2\textwidth}
        \includegraphics[width=\textwidth]{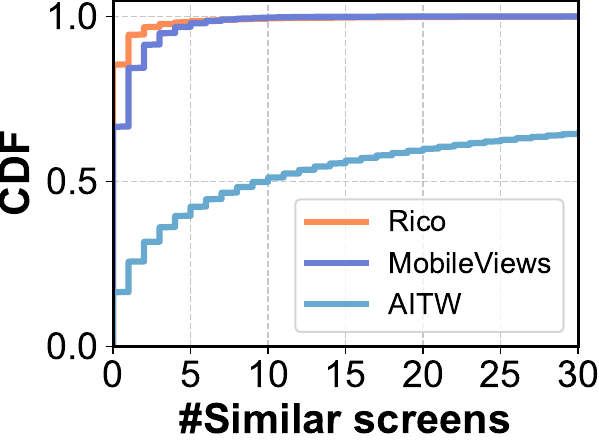}
        \caption{}
        \label{fig:cdf-similar-ui}
    \end{subfigure}
    \vspace{-4pt}
    \caption{Screen diversity analysis. (a) Distribution of the number of screens collected per app in \sys. (b) Cumulative distribution function (CDF) of similar screens in Rico, AITW, and \sys, where the X-axis denotes the number of similar screens corresponding to each screen in a dataset.}
    \vspace{-13pt}
    \label{fig:ui-dist-cdf}
\end{figure}

\textbf{Screen diversity analysis.}
We examine \sys's screen diversity from two perspectives:
(1) the distribution of screens collected per app, and
(2) screen similarity across large-scale mobile GUI datasets.
The second perspective is motivated by the fact that, before image de-duplication, AITW contains over 5M screens from only 357 apps, suggesting a considerable amount of visually similar content.
To measure similarity, we use \textit{ImageHash}~\cite{imagehash} to compute a perceptual hash for each screen, and treat two screens as similar if their hashes are identical.
Below, we examine these two aspects through quantitative analysis.

$\bullet$ \textit{How many screens are collected per app in \sys?}
Figure~\ref{fig:ui-cnt-distribution} shows the distribution of screens collected per app.
Most apps contribute 10--20 unique screens, reflecting the natural variation in app complexity: simple apps (e.g., calculators) expose fewer screens, whereas complicated apps (e.g., social media) expose more.
Notably, 6.8\% of apps yield over 100 screens, indicating that our VLM-enhanced traversal is able to explore apps extensively.
Overall, this distribution demonstrate that \sys achieves broad app coverage while ensuring GUI coverage within apps.

$\bullet$ \textit{How does screen uniqueness compare across datasets?}
Figure~\ref{fig:cdf-similar-ui} compares screen similarity across Rico~\cite{deka2017rico}, AITW~\cite{rawles2023android}, and \sys.
For each screen, we count how many visually similar screens appear in the same dataset and use the cumulative distribution function (CDF) to summarize these counts.
Both \sys and Rico exhibit high screen uniqueness: over 95\% of screens have fewer than 5 similar counterparts, and the curves saturate rapidly due to their broad app coverage.
In contrast, AITW shows severe redundancy.
The CDF rises slowly and plateaus around 0.7, meaning 30\% of AITW's screens have more than 30 similar duplicates.
More specifically, over 1.2M screens (24\% of the dataset) have over 100 similar counterparts, and the most duplicated screen appears 67K times, accounting for 1.3\% of the dataset.
These comparisons show that \sys balances scale and diversity, providing a dataset suitable for training mobile GUI agents that require broad perceptual coverage and stronger generalization.

%% file: sections/sec-experiment.tex
\section{Experiments}
\label{sec:experiment}
As described in $\S$\ref{sec:dataset-preview}, we select 13K representative screens from \sys and construct 26K GUI grounding samples through multi-stage filtering and VLM-assisted annotation.
To assess the data quality of \sys, we train four VLMs on these samples and evaluate their performance on established GUI grounding benchmarks.
In this section, we first describe the training algorithm and experimental setup.
We then (1) compare our trained VLMs with state-of-the-art models, and (2) investigate how training data filtering strategy and data scale affect VLM performance.

\subsection{Training Algorithm}
In realistic GUI interactions, a click is considered successful when the predicted point falls within the target GUI component.
However, supervised fine-tuning (SFT) enforces an exact token-level match between the model output and the ground truth, which diverges from this practical interaction criterion.
To better align training objectives with real-world execution, we adopt the Group Relative Policy Optimization (GRPO) algorithm, optimized in DeepSeek-R1~\cite{deepseekr1}.
GRPO removes the need for an additional reward model by using structured, verifiable rewards, which substantially reduces training overhead while allowing the reward signal to be directly tied to whether the predicted point hits the correct GUI area.
Several recent studies~\cite{gu2025uivenus,lu2025uir1,liu2025infiguir1,luo2025guir1,gao2025uishift} further validate the effectiveness of GRPO by introducing customized reward functions tailored to GUI tasks.
In this study, we employ GRPO with a task-specific reward function that reflects the hit accuracy of predicted points in GUI grounding.

\textbf{Training objectives.} 
Following DeepSeek-R1~\cite{deepseekr1}, our training objective aims to maximize the expected clipped advantage reward while penalizing divergence from a reference policy.
Specifically, given a GUI grounding question $q$, we sample $G$ responses $\{o_i\}_{i=1}^G$ from the old policy $\pi_{\theta_{old}}$.
For each sample, we compute the probability ratio $\rho_i = \frac{\pi_\theta(o_i \mid q)}{\pi_{\theta_{old}}(o_i \mid q)}$, and the advantage $A_i = \frac{r_i - \mathrm{mean}(\{r_1, \ldots, r_G\})}{\mathrm{std}(\{r_1, \ldots, r_G\})}$, where $r_i$ is the reward assigned to the $i$-th response based on our designed reward function.
$\epsilon$ is the clipping parameter to stabilize training, $\beta$ is the KL penalty coefficient, and $\pi_{\mathrm{ref}}$ is the reference policy used for KL divergence calculation.
The overall GRPO objective is defined as:
\begin{align}
    \mathcal{J}_{\mathrm{GRPO}}(\theta)
    &= \mathbb{E}\left[ q \sim P(Q), \{o_i\}_{i=1}^G \sim \pi_{\theta_{old}}(O \mid q) \right] \nonumber \\
    &\quad \frac{1}{G} \sum_{i=1}^G \Bigg[
        \min\left(
            \rho_i A_i,\,
            \mathrm{clip}\left(
                \rho_i,\,
                1 - \epsilon,\,
                1 + \epsilon
            \right) A_i
        \right) \nonumber \\
    &\qquad\qquad - \beta\, \mathbb{D}_{\mathrm{KL}}(\pi_\theta \| \pi_{\mathrm{ref}})
    \Bigg]
    \label{eq:grpo_obj}
\end{align}

\textbf{Reward design for GUI grounding.} 
In GUI grounding, a click is deemed successful if the predicted point lies within the target GUI component's bounding box.
Accordingly, we design a straightforward reward function that assigns a reward of 1 if the predicted point falls within the ground truth bounding box, and 0 otherwise:
\begin{equation}
    R =
    \begin{cases}
        1, & \text{if point in bounding box} \\
        0, & \text{otherwise}
    \end{cases}
    \label{eq:reward_func}
\end{equation} 
This binary reward effectively guides the model to learn accurate grounding strategies aligned with practical interaction requirements.
Different from previous studies~\cite{deepseekr1,gu2025uivenus,lu2025uir1,liu2025infiguir1,luo2025guir1,gao2025uishift} that include a format reward to enforce output consistency and facilitate result parsing, 
we observe that, when using the official GUI grounding prompts provided for each model, the models are able to produce the required output format consistently from the start of training.
Therefore, we do not include the external format correctness reward and instead require the model to predict the center point directly, enabling it to focus exclusively on the grounding objective.

\subsection{Experimental Setup}
\textbf{Training data selection.} 
To validate the effectiveness of \sys in enhancing GUI grounding capabilities and to investigate how data quality and scale affect model performance, we design two comparative settings:
\begin{itemize}
    \item \textit{Random scaling.} To examine the impact of data scale, we randomly sample subsets of different sizes from the constructed 26K dataset for training and analyze performance trends as the scale increases.
    \item \textit{Model-specific filtering.} To assess the effect of data quality, we perform model-specific data filtering before GRPO training to select samples better aligned with each model's learning capacity.
    Specifically, for each GUI grounding sample, we use the base model to generate $G=8$ responses and score them with the reward function defined in Eq.~\ref{eq:reward_func}, producing a total score between 0 and 8.
    Samples with scores of 0 or 8 are discarded, as they represent either overly difficult or trivial cases that provide limited learning value, while samples with intermediate scores (1--7) are retained for subsequent training.
\end{itemize}

\textbf{Training setup.}
To evaluate the impact of \sys on both general-purpose and GUI-specific VLMs, we select four models: Qwen2.5-VL-7B~\cite{bai2025qwen25vltechnicalreport}, MiMo-VL-7B-SFT~\cite{coreteam2025mimovltechnicalreport}, MiMo-VL-7B-RL~\cite{coreteam2025mimovltechnicalreport}, and UI-Tars-1.5~\cite{qin2025uitars}.
All models participate in the random data scaling experiments.
For the model-specific data filtering experiments, we apply the filtering strategy individually to each model and obtain filtered subsets containing 1K, 1.3K, and 254 samples for MiMo-VL-7B-SFT, MiMo-VL-7B-RL, and UI-Tars-1.5, respectively.
For Qwen2.5-VL-7B, we observe that all samples receive either 0 or 8 reward scores when using a high sampling temperature; therefore, this model is only included in the data scaling experiments.
We implement the GRPO training based on the VLM-R1~\cite{shen2025vlmr1} framework with the vision encoder frozen and the language model fully fine-tuned.
We train each model for four epochs with sampling temperature 0.9, KL penalty $\beta=0.04$, and eight responses per sample.
All experiments are conducted on 8 $\times$ H100 GPUs.

\textbf{Benchmarks and metrics.}
We evaluate all models on two established GUI grounding benchmarks: ScreenSpot-V2~\cite{wu2024atlas}, a refined version of ScreenSpot~\cite{cheng2024seeclick} with corrected annotations, containing 1,272 samples from mobile, desktop, and web platforms; and ScreenSpot-Pro~\cite{li2025screenspotproguigroundingprofessional}, a more challenging benchmark comprising 1,581 samples focusing on high-resolution grounding.
We adopt center-point accuracy as the evaluation metric, which measures the proportion of predicted points that fall within the ground-truth bounding boxes of the target GUI components.

\subsection{Results and Analysis}
We first validate the effectiveness of \sys in improving GUI grounding performance, then analyze how data quality and scale impact model performance.

\textbf{Comparison with state-of-the-art models.}
Table~\ref{tab:gui-grounding} summarizes the results on ScreenSpot-v2 and ScreenSpot-Pro for performant VLMs and four models trained with \sys.
In general, models trained with \sys show clear improvements over their original versions and perform competitively with prior strong models.
On ScreenSpot-v2, MiMo-VL-7B-SFT~\cite{coreteam2025mimovltechnicalreport} trained with \sys surpasses UI-Venus-Ground-7B~\cite{gu2025uivenus} in both mobile and web text grounding, achieving an average accuracy of 91.0\%.
On ScreenSpot-Pro, which requires high-resolution grounding, all four models trained with \sys achieve consistent improvements ranging from 2.0\% to 4.5\%.
Notably, UI-Tars-1.5~\cite{qin2025uitars} improves from 89.0\% to 90.3\% on ScreenSpot-v2 and from 41.6\% to 46.1\% on ScreenSpot-Pro, indicating that \sys benefits models already optimized for GUI tasks.
These results demonstrate that \sys consistently enhances GUI grounding across both general-purpose and GUI-specific VLMs.

\input{tables/gui-grounding.tex}
\input{tables/gui-grounding-filter.tex}
\textbf{Effects of model-specific training data filtering.}
To isolate the impact of filtering itself, we train each model with an equal number of unfiltered samples and directly compare their performance with the filtered counterparts.
As shown in Table~\ref{tab:gui-grounding-filter}, training with unfiltered \sys data already improves all models over their baselines, indicating the inherent quality of the \sys datatset.
Applying model-specific filtering yields further gains: on ScreenSpot-v2, improvements range from 1.3\% to 3.4\%, while on ScreenSpot-Pro, gains span 1.4\% to 3.3\%.
For general-purpose models, MiMo-VL-7B-SFT and MiMo-VL-7B-RL, filtering removes noisy samples and stabilizes optimization, leading to consistent performance increases across both benchmarks.
For the GUI-specific model UI-Tars-1.5, which is trained with only 254 samples, filtering still yields over a 1\% improvement on both benchmarks, demonstrating its effectiveness under limited training data.
These results confirm that \sys inherently provides effective data for training, while filtering further amplifies these benefits, enabling more efficient and robust model improvement.

\input{figs/gui-grounding-scale.tex}
\textbf{Impact of training data filtering strategy and data scale.}
Figure~\ref{fig:gui-gronding-scale} compares the results of random data scaling and model-specific data filtering across four VLMs.
For general-purpose models, performance under random scaling tends to fluctuate or even decline as the training data increases, suggesting that simply enlarging the dataset introduces redundancy and noise.
In contrast, model-specific filtering yields more stable improvements of 0.9--3.3\% on both benchmarks for all models, showing that aligning data selection with each model's learning capacity provides more effective supervision.
On SreenSpot-v2, UI-Tars-1.5 trained with only 254 filtered samples performs comparably to that trained with 26K random samples, confirming that targeted filtering can match or exceed the benefits of large-scale random data.
These results demonstrate that model-specific training data filtering is more effective than simply scaling training data with random samples.
Such filtering is enabled by the large scale and diversity of \sys, which provides a rich pool for extracting model-specific samples tailored to different VLMs' learning capacities.

%% file: tables/gui-grounding.tex
\begin{table}[t]
    \centering
    \caption{
      Performance comparison on GUI grounding benchmarks: ScreenSpot-v2 and ScreenSpot-Pro.
      Training with \sys enhances grounding ability across both general-purpose and GUI-specific VLMs.
      \textbf{Bold}: best result; \underline{underlined}: second best.
    }
    \setlength{\tabcolsep}{4pt}
    \label{tab:gui-grounding}
    \vspace{-4pt}
    \scalebox{0.71}{%
    \begin{tabular}{@{}
    >{\columncolor[HTML]{FFFFFF}}l 
    >{\columncolor[HTML]{FFFFFF}}c 
    >{\columncolor[HTML]{FFFFFF}}c 
    >{\columncolor[HTML]{FFFFFF}}c 
    >{\columncolor[HTML]{FFFFFF}}c 
    >{\columncolor[HTML]{FFFFFF}}c 
    >{\columncolor[HTML]{FFFFFF}}c 
    >{\columncolor[HTML]{FFFFFF}}c 
    >{\columncolor[HTML]{FFFFFF}}c @{}}
    \toprule
    \cellcolor[HTML]{FFFFFF} &
      \multicolumn{7}{c}{\cellcolor[HTML]{FFFFFF}\textbf{ScreenSpot-v2}} &
      \textbf{-Pro} \\ \cmidrule(lr){2-8} \cmidrule(l){9-9}
    \cellcolor[HTML]{FFFFFF} &
      \multicolumn{2}{c}{\cellcolor[HTML]{FFFFFF}Mobile} &
      \multicolumn{2}{c}{\cellcolor[HTML]{FFFFFF}Desktop} &
      \multicolumn{2}{c}{\cellcolor[HTML]{FFFFFF}Web} &
      \cellcolor[HTML]{FFFFFF} &
      \cellcolor[HTML]{FFFFFF} \\ \cmidrule(l){2-3} \cmidrule(l){4-5} \cmidrule(l){6-7}
    \multirow{-3}{*}{\cellcolor[HTML]{FFFFFF}\textbf{Model}} &
      Text &
      Icon &
      Text &
      Icon &
      Text &
      Icon &
      \multirow{-2}{*}{\cellcolor[HTML]{FFFFFF}Avg.} &
      \multirow{-2}{*}{\cellcolor[HTML]{FFFFFF}Avg.} \\ \midrule
    CogAgent-18B~\citep{hong2023cogagent}         & -             & -             & -             & -             & -             & -             & -             & 7.7               \\
    SeeClick-9.6B~\citep{cheng2024seeclick}        & 78.4          & 50.7          & 70.1          & 29.3          & 55.2          & 32.5          & 55.1          & 1.1               \\
    UGround-7B~\citep{gou2025uground}         & 75.1          & 84.5          & 85.1          & 61.4          & 84.6          & 71.9          & 76.3          & 16.5              \\
    OS-Atlas-7B~\citep{wu2024atlas}          & 95.2          & 75.8          & 90.7          & 63.6          & 90.6          & 77.3          & 84.1          & 18.9              \\
    ShowUI-2B~\citep{lin2024showui}            & -             & -             & -             & -             & -             & -             & -             & 7.7               \\
    UI-R1-E-3B~\citep{lu2025uir1}           & 83.0          & \textbf{97.1} & 85.0          & \textbf{91.7} & 77.9          & \textbf{95.4} & 89.5          & 33.5              \\
    InfiGUI-R1-3B~\citep{liu2025infiguir1}        & -             & -             & -             & -             & -             & -             & -             & 35.7              \\
    TongUI-7B~\citep{zhang2025tongui}   & 93.1          & 81.5    &  96.4          & 82.9    & 90.2 & 84.7    &  88.7 & 24.7     \\ 
    UI-Venus-Ground-7B~\citep{gu2025uivenus}   & {\ul 99.0}          & {\ul 90.0}    & \textbf{97.0}          & {\ul 90.7}    & {\ul 96.2} & {\ul 88.7}    & \textbf{94.1} & \textbf{50.8}     \\\midrule
    \multicolumn{9}{c}{\cellcolor[HTML]{F3F3F3}\textit{Ours: Four models trained with \sys}}                                                                     \\
    Qwen2.5-VL-7B~\citep{bai2025qwen25vltechnicalreport}        & 98.3    & 86.3          & 88.7          & 67.1          & 92.7          & 81.8          & 87.7          & 26.4              \\
    \textit{ w/ \sys}    & {\ul 99.0}          & 84.4    & 89.2          & 75.7          & 94.4          & 81.8          & \cellcolor[HTML]{D4F6FF}89.2\deltapos{1.5}   & \cellcolor[HTML]{D4F6FF}28.4\deltapos{2.0}       \\ \midrule
    Mimo-VL-7B-SFT~\citep{coreteam2025mimovltechnicalreport}       & 96.6          & 84.4          & 92.8          & 80.0          & 88.9          & 76.8          & 87.6          & 39.8              \\
    \textit{ w/ \sys} & \textbf{99.3} & 85.3 & 93.3 & 81.4    & \textbf{96.6}   & 83.3 & \cellcolor[HTML]{88d8db}{\ul 91.0}\deltapos{3.4} & \cellcolor[HTML]{D4F6FF}42.1\deltapos{2.3}\\\midrule
    Mimo-VL-7B-RL~\citep{coreteam2025mimovltechnicalreport}        & 98.3    & 86.3          & 90.2          & 80.7          & 92.7          & 75.4          & 88.4          & 40.2              \\
    \textit{ w/ \sys}  & \textbf{99.3}   & 88.2 & 94.3 & 82.1   & 92.3 & 76.4 & \cellcolor[HTML]{D4F6FF}89.9\deltapos{1.5}& \cellcolor[HTML]{D4F6FF}43.5\deltapos{3.3} \\\midrule
    UI-Tars-1.5~\citep{qin2025uitars}        & 95.5   & 84.4          & 91.2          & 78.6          & 91.5          & 86.7          & 89.0          & 41.6              \\
    \textit{ w/ \sys}  & 98.3   & 82.9 & 94.3 & 79.3   & 93.6 & 86.7 & \cellcolor[HTML]{D4F6FF}90.3\deltapos{1.3}& \cellcolor[HTML]{88d8db}{\ul 46.1}\deltapos{4.5} \\\bottomrule
    \end{tabular}
    }
    \vspace{-2pt}
\end{table}

%% file: tables/gui-grounding-filter.tex
\begin{table}[t]
    \centering
    \caption{
      Performance comparison on GUI grounding benchmarks: ScreenSpot-v2 and ScreenSpot-Pro.
    Each model is trained with the same number of samples, using either unfiltered or filtered data from \sys.
    Filtering improves data quality and leads to consistent performance gains across all models.
    }
    \setlength{\tabcolsep}{4pt}
    \label{tab:gui-grounding-filter}
    \vspace{-4pt}
    \scalebox{0.71}{%
    \begin{tabular}{@{}
    >{\columncolor[HTML]{FFFFFF}}l 
    >{\columncolor[HTML]{FFFFFF}}c 
    >{\columncolor[HTML]{FFFFFF}}c 
    >{\columncolor[HTML]{FFFFFF}}c 
    >{\columncolor[HTML]{FFFFFF}}c 
    >{\columncolor[HTML]{FFFFFF}}c 
    >{\columncolor[HTML]{FFFFFF}}c 
    >{\columncolor[HTML]{FFFFFF}}c 
    >{\columncolor[HTML]{FFFFFF}}c @{}}
    \toprule
    \cellcolor[HTML]{FFFFFF} &
      \multicolumn{7}{c}{\cellcolor[HTML]{FFFFFF}\textbf{ScreenSpot-v2}} &
      \textbf{-Pro} \\ \cmidrule(lr){2-8} \cmidrule(l){9-9}
    \cellcolor[HTML]{FFFFFF} &
      \multicolumn{2}{c}{\cellcolor[HTML]{FFFFFF}Mobile} &
      \multicolumn{2}{c}{\cellcolor[HTML]{FFFFFF}Desktop} &
      \multicolumn{2}{c}{\cellcolor[HTML]{FFFFFF}Web} &
      \cellcolor[HTML]{FFFFFF} &
      \cellcolor[HTML]{FFFFFF} \\ \cmidrule(l){2-3} \cmidrule(l){4-5} \cmidrule(l){6-7}
    \multirow{-3}{*}{\cellcolor[HTML]{FFFFFF}\textbf{Model}} &
      Text &
      Icon &
      Text &
      Icon &
      Text &
      Icon &
      \multirow{-2}{*}{\cellcolor[HTML]{FFFFFF}Avg.} &
      \multirow{-2}{*}{\cellcolor[HTML]{FFFFFF}Avg.} \\ \midrule
    \multicolumn{9}{c}{\cellcolor[HTML]{F3F3F3}\textit{Mimo-VL-7B-SFT trained with 1K samples}}\\                                                                 
    Mimo-VL-7B-SFT~\cite{coreteam2025mimovltechnicalreport}       & 96.6          & 84.4          & 92.8          & 80.0          & 88.9          & 76.8          & 87.6          & 39.8              \\
    \textit{ w/o data filtering} & 99.0 & 86.7 & 92.3 & 79.3    & 91.5   & 76.8 & 88.8\deltapos{1.2} & 40.7\deltapos{0.9}\\
    \rowcolor[HTML]{D4F6FF}
    \textbf{\textit{ w/ data filtering}} & 99.3 & 85.3 & 93.3 & 81.4    & 96.6   & 83.3 & \textbf{91.0\deltapos{3.4}} & \textbf{42.1\deltapos{2.3}}\\\midrule
    \multicolumn{9}{c}{\cellcolor[HTML]{F3F3F3}\textit{Mimo-VL-7B-RL trained with 1.3K samples}}\\ 
    Mimo-VL-7B-RL~\cite{coreteam2025mimovltechnicalreport}        & 98.3    & 86.3          & 90.2          & 80.7          & 92.7          & 75.4          & 88.4          & 40.2              \\
    \textit{ w/o data filtering}  & 99.3   & 86.7 & 92.3 & 80.7   & 93.6 & 76.8 & 89.5\deltapos{1.1}& 41.6\deltapos{1.4} \\
    \rowcolor[HTML]{D4F6FF}
    \textbf{\textit{ w/ data filtering}}  & 99.3   & 88.2 & 94.3 & 82.1   & 92.3 & 76.4 & \textbf{89.9\deltapos{1.5}}& \textbf{43.5\deltapos{3.3}} \\\midrule
    \multicolumn{9}{c}{\cellcolor[HTML]{F3F3F3}\textit{UI-Tars-1.5 trained with 254 samples}} \\
    UI-Tars-1.5~\cite{qin2025uitars}        & 95.5   & 84.4          & 91.2          & 78.6          & 91.5          & 86.7          & 89.0          & 41.6              \\
    \textit{ w/o data filtering}  & 96.2   & 83.9 & 92.8 & 79.3   & 91.5 & 87.2 & 89.5\deltapos{0.5}& 41.7\deltapos{0.1} \\
    \rowcolor[HTML]{D4F6FF}
    \textbf{\textit{ w/ data filtering}}  & 97.2   & 83.9 & 93.3 & 82.1   & 92.7 & 86.7 & \textbf{90.3\deltapos{1.3}}& \textbf{43.0\deltapos{1.4}} \\\bottomrule
    \end{tabular}
    }
    \vspace{-13pt}
\end{table}

%% file: figs/gui-grounding-scale.tex
\begin{figure}[t]
	\centering	
	\vspace{-5pt}	
	\includegraphics[width=0.49\textwidth]{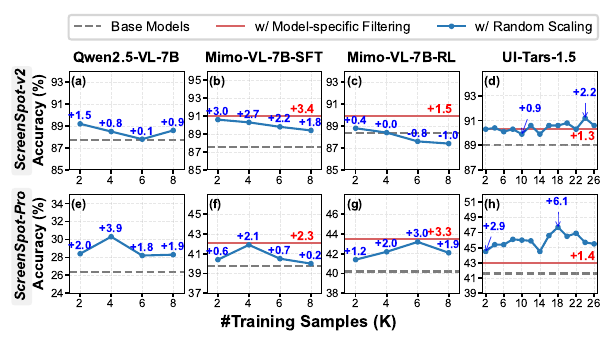}
	\vspace{-20pt}				
	\caption{
		Effects of random data scaling and model-specific training data filtering on GUI grounding performance across models.
    }
    \vspace{-15pt}
	\label{fig:gui-gronding-scale}
\end{figure}

%% file: sections/sec-conclusion.tex
\section{Conclusion}
We develop \sys Crawler, a scalable mobile app traversal framework for large-scale data collection.
Building on this, we construct \sys, an open-source dataset of 1.2M unique screenshot-VH pairs from over 30K modern apps with comprehensive metadata and bilingual support.
Experiments show that \sys consistently improves both general-purpose and GUI-specific VLMs. 
Model-aware training data filtering further enhancing training efficiency.
We believe that \sys can serve as a reliable foundation for advancing mobile GUI agents.